\documentclass[preprint,authoryear,12pt]{elsarticle}

\usepackage[utf8]{inputenc} 
\usepackage{geometry} 
\usepackage{graphicx} 

\usepackage{array} 
\usepackage{verbatim} 
\usepackage{fancyhdr} 
\usepackage{amsmath}
\usepackage{amsfonts}
\usepackage{amssymb}
\usepackage{amsthm}
\usepackage{float}
\usepackage[lined,algonl,boxed]{algorithm2e}
\usepackage[table]{xcolor}
\usepackage{caption}
\usepackage{natbib}

\newtheorem{example}{Example}
\newtheorem{theorem}{Theorem}
\newtheorem{lemma}{Lemma}
\newtheorem{corollary}{Corollary}
\newtheorem{definition}{Definition}

\usepackage{todonotes}

\newcommand{\uprom}[1]{\uppercase\expandafter{\romannumeral#1}}

\journal{ }

\begin{document}

\begin{frontmatter}



\title{Truthfulness with Value-Maximizing Bidders}


\author{Salman Fadaei and Martin Bichler*}

\address{Department of Informatics, Technical University of Munich, Germany \\
Email: salman.fadaei@gmail.com, bichler@in.tum.de. }
\cortext[cor1]{Corresponding author: Martin Bichler, Department of Informatics, Technical University of Munich, Boltzmannstr. 3, 85748 Garching, Germany}
\fntext[label2]{The financial support from the Deutsche Forschungsgemeinschaft
(DFG) (BI 1057/1-4) is gratefully acknowledged.}
\fntext[label3]{A preliminary version of this paper has been accepted for publication in the proceedings of the 9th International Symposium on Algorithmic Game Theory (SAGT 2016).
}

\begin{abstract}
In markets such as digital advertising auctions, bidders want to maximize value rather than payoff. This is different to the utility functions typically assumed in auction theory and leads to different strategies and outcomes. We refer to bidders who maximize value as \textit{value bidders}. 
While simple single-object auction formats are truthful, standard multi-object auction formats allow for manipulation. It is straightforward to show that there cannot be a truthful and revenue-maximizing deterministic auction mechanism with value bidders and general valuations. Approximation has been used as a means to achieve truthfulness, and we study which approximation ratios we can get from truthful approximation mechanisms. We show that the approximation ratio that can be achieved with a deterministic and truthful approximation mechanism with $n$ bidders and $m$ items cannot be higher than $1/n$ for general valuations. For randomized approximation mechanisms there is a framework with a ratio of $O(\frac{\sqrt{m}}{\epsilon^3})$ with probability at least $1-\epsilon$, for $0<\epsilon<1$. We provide better ratios for environments with restricted valuations.
\end{abstract}


\end{frontmatter}


\section{Introduction}
Auctions have received increasing attention in operations research for the coordination of supply chains \citep{Lorentziadis2016347, Mansouri2015565}.
In auction theory, bidders are almost always modeled as payoff-maximizing individuals using a quasilinear utility function. 
Under these utility functions the Vickrey-Clarke-Groves mechanism is the unique mechanism to obtain maximum welfare in dominant strategies. 
However, there are markets where a pure quasilinear utility function might just not be the right assumption. Sometimes, capacity constraints \citep{Chaturvedi2015987} or budget constraints \citep{Dobzinski12} need to be considered which has ample effect on the equilibrium bidding strategies. Sometimes, however, payoff-maximization might just not be the right assumption and bidders rather maximize value subject to a budget constraint.

For example, digital advertising markets have grown substantially in the recent years \citep{Ember15}. 
In \textit{display ad auctions} individual user impressions on a web site are auctioned off. Advertising buyers bid on an impression 
and, if the bid is won, the buyer's ad is instantly displayed on the publisher’s site. 
Demand-side platforms (DSPs) are intermediaries, who provide the technology to bid for advertisers 
on such advertising exchanges. 
A number of papers describe bidding strategies and heuristics in display ad auctions 
\citep{Feldman08, ghosh2009adaptive, chen2011real, Elmeleegy2013DMPTurn, Zhang2014OptimalRTBUCL, zhang2016feedback}. 
\cite{Zhang2014OptimalRTBUCL} gives an up-to-date overview.  
In all of these papers the task of the DSP or advertiser is to maximize the values of impressions typically subject to a budget constraint for a campaign.


Value maximization subject to a budget is not limited to display ad auctions. Private individuals often determine a budget before making a purchase, and then buy the best item or set of items (e.g., cars, real-estate) that meets the budget. Actually, in classical micro-economic consumer choice theory, consumers select a package of objects that maximizes value subject to a budget constraint, they don't maximize payoff \citep{mas1995microeconomic}. Maximizing value subject to a budget constraint is also wide-spread in business due to principal-agent relationships \citep{engelbrecht1987optimal}. For example, in spectrum auctions, national telecoms have different preferences for different packages of spectrum licenses based on the corresponding net present values of business cases. These billion dollar net present values exceed the financial capabilities of the local telecom by far, but not those of its stakeholder, a multinational, which has mainly long run strategic incentives of operating in the local market. Thus, the stakeholder provides the local telecom with allowances for individual packages based on the underlying net present value. It is well-known by practitioners that there often is a principal-agent relationship between the shareholder and the bidding team \citep{Shapiro13}. Both parties determine the valuations of different allocations and the principal then sets allowances less or equal to these valuations, which describe how high the agent can bid in order to win a package. \citet{Shapiro13} argue that such pre-determined budgets have to do with capital rationing \citep{Paik95}. For agents, these allowances are like sunk costs but they have preferences to win the most valuable packages within budget.


In this paper, we analyze truthful approximation mechanisms for value maximizing bidders. One can think of several market types with value bidders depending on the nature of the preferences for one or more objects and of the budgets:

\begin{itemize}
	\item Single-object markets, where multiple value bidders compete for only one object
	\item Assignment markets, where each bidder is interested to win at most one out of many objects
	\item Generalized assignment markets, where bidders want to win one or more objects, they have values for each object, and an overall budget constraint
	\item Combinatorial markets, where bidders have preferences for and can bid on packages of objects
\end{itemize}

In display ad auctions budgets for an advertising campaign are large compared to the cost of an impression such that simple heuristics for online knapsack problems allow for near-optimal solutions \citep{Borgs07, zhou2008budget}. In such markets, an advertiser can consider individual impressions separately and bid on any impression which promises a certain return on investment, i.e., a certain ratio of expected payoff to price of an impression. We can deal with each impression individually and focus on the model of single-object markets with value bidders. A different model is that where the budget constraint is binding and a buyer wants to buy as many objects (or impressions) as possible within budget. Such an environment could be described as generalized assignment market. Assignment markets, where each buyer wants to win at most one object are in between. Finally, buyers might have preferences for packages of objects such as in the spectrum auction markets in our introduction, i.e., we face a combinatorial market. We discuss truthful mechanisms for a number of these environments, but focus on approximation mechanisms for combinatorial auction markets (with a fully expressive XOR bid language). Combinatorial auction markets allow for general preferences including substitutes and complements and the efficiency of the mechanism is bounded just due to restrictions in the expressiveness of the bid language.

\subsection{Our results}

First, we analyze a truthful Pareto-optimal mechanism for markets with value bidders. We show that such a mechanism exists. 
Then we study truthful revenue maximizing mechanisms. We focus on revenue rather than welfare. Social welfare is difficult to analyze in environments where bidders have values and budget constraints.\footnote{Consider the case of two bidders, one with a high value and low budget for an object, and another one with a lower value and a high budget constraint. In a revenue-maximizing auction, we only need to consider the willingness-to-pay for the object including the available budget and aim for the allocation that maximizes the auctioneers' revenue. } We show that for single-minded and single-valued value bidders there are simple truthful mechanisms that maximize revenue, 
but that this is not possible for multi-minded value bidders. Next, we explore truthful approximation mechanisms. 

The need for approximating revenue arises for two reasons. One is because the underlying optimization problem is computationally intractable. 
This has been the primary motivation for approximation mechanisms as they are described in \cite{Nisan07}. 
In contrast, we approximate revenue to obtain truthfulness. The approach is similar to that of \citet{Procaccia09}: 
We maximize revenue without considering incentives, and refer to this as optimal revenue. We will then say that a strategy-proof mechanism returns (at least) a ratio $\alpha$ of the optimal if it's revenue is always greater than or equal to $\alpha$ times the optimal revenue.

We first look at small markets with two bidders and two items only to get some intuition about possibilities for manipulation, and find out that with a simple assumption there is a truthful deterministic mechanism with a golden approximation factor of $\frac{\sqrt 5 - 1}{2}$. 
A randomized mechanism for the environment with two bidders and two units achieves a factor of $\frac{3}{4}$. 
Unfortunately, the deterministic mechanism cannot be extended to larger markets. 
The analysis of these small markets is valuable on its own right, but it is also helpful for deriving our main result:


\textbf{Theorem.} \textit{The best revenue ratio achievable by a deterministic and truthful mechanism with value bidders in a market with $n$ bidders and $m$ homogeneous items is $\frac{1}{n}$, for any $n \geq 2$ and $m\geq 2$.}

The theorem has a straightforward extension to combinatorial markets. 
In quasi-linear mechanism design, randomization is often a remedy to achieve higher approximation ratios. 
Approximation mechanisms for quasi-linear bidders do typically not lead to strategy-proofness with value bidders. 
However, there is a recent contribution by \citet{Dobzinski12random}, which is also truthful for value bidders with a simple change of the payment rule. 

\textbf{Theorem.} \textit{There exists a polynomial-time randomized mechanism for value bidders which is universally truthful and guarantees an approximation ratio of $O(\frac{\sqrt{m}}{\epsilon^3})$ with probability at least $1-\epsilon$, for $0<\epsilon<1$.}


\subsection{Related literature}

Given the substantial literature in social choice, we first position \textit{value bidders} in the literature. 
The Gibbard-Satterthwaite theorem describes one of the most celebrated results in social choice theory. 
\citet{Gibbard73} proved that any non-dictatorial voting scheme with at least three possible outcomes is not strategy-proof. 
\citet{Satterthwaite75} showed that if a committee is choosing among at least three alternatives, then every strategy-proof voting procedure is dictatorial. 
There have been a number of extensions analyzing more specific mechanism design questions for allocation problems without money, typically resulting in impossibility results \citep{Papai01, Ehlers03, Hatfield09}. 


\textit{Quasi-linear preferences} are an escape route from the impossibilities found above. The well-known result by \citet{Green79} shows that the VCG mechanism is the unique quasi-linear mechanism, which allows for strategy-proofness and efficiency. 
There is a huge literature on approximation mechanisms for quasi-linear bidders. 
The computational hardness for the algorithmic problem of revenue maximization with general valuations in  combinatorial auctions is shown to be $O(\sqrt{m})$ \citep{Halldorsson00}.
This is a natural lower bound on the approximation factor of truthful approximation mechanisms. For quasi-linear bidders randomized approximation mechanisms with the same approximation ratio have been found \citep{Lavi11, Dobzinski12random}. However, the best deterministic truthful approximation guarantee known for general combinatorial auctions is $O(\frac{m}{\sqrt{\log m}})$ \citep{Holzman04}.

Closest to our assumptions is the model analyzed by \citet{Feldman08} in which bidders have an overall budget and a value for ad slots in sponsored search and they want to maximize the number of clicks given their budget. They also argue that a bidder is incentivized to spend the entire budget to maximize exposure or the number of clicks in the market. 
\citet{Feldman08} focused on the specifics of ad slot markets with purely additive valuations for homogeneous goods (clicks) and they consider specific scheduling constraints. The overall budget can be seen as the budget for packages in which clicks outnumber a specific threshold. In our model, we do not restrict valuations to be additive and we are interested in packages of heterogeneous items and different budgets for different packages. Recently, we became aware of a working paper by \cite{wilkens2016}, who also motivates value maximization, but the paper has a different focus.


\subsection{Paper structure}
In Section 2 we introduce necessary notation and definitions used throughout the paper. 
In Section 3 we present a Pareto-optimal mechanism and prove that revenue maximization and strategyproofness in general are impossible. 
Section 4 focuses on approximation mechanisms in markets with two units of a good and two bidders only, where we achieve good approximation ratios. 
In Section 5 we show that for general markets without restrictions on the valuations no strategyproof approximation mechanism can achieve a better approximation ratio than $\frac{1}{n}$. 
In Section \ref{general-randomized} we introduce a randomized approximation mechanism for value bidders with general valuations that is universally truthful.
Finally, we conclude with a summary and a discussion about future research questions.

\section{Preliminaries and notations} \label{model_intro}

In a combinatorial market we have $m$ non-homogeneous items, $J$, one seller $0$, and $n$ bidders, $I$ ($I_0$ includes the seller). Each bidder $i \in I$ has a valuation $v_i(a_i)$ for any package $a_i \subseteq a$ assigned to an agent $i$ and allocation $a \in A$, where $A$ describes the set of all allocations.  
A feasible allocation of bundles of items to bidders is described as $a = \bigcup_{i \in I}a_i$ with $\bigcap_{i \in I}a_i=\emptyset$ and $a \in A$. For brevity, we will drop the subscript in $a_i$ and write $v_i(a)$, even though the bidder is only interested in his own allocation and not the allocation overall.
When we discuss combinatorial auctions with heterogeneous items, we assume general cardinal valuations and allow for substitutes and complements and free disposal ($v_i(S) \leq v_i(T)$, $S \subseteq T\subseteq J$). In contrast to mechanism design with quasi-linear utility functions where utility is defined as valuation minus price of a bundle, $u_i(a)=v_i(a)-p_i(a)$, we assume that these bidders have no value for residual budget or payoff, but they want to win their highest-valued package subject to some budget constraint $b_i$. 

There are different ways, how budget can be considered. In combinatorial markets the value of larger packages of objects $a$ can exceed the overall budget constraint $v_i(a) > b_i(a)$. This can also happen in single object auctions. In this paper we are concerned with maximizing revenue, and we consider the willingness-to-pay for a package rather than the true value and trim the valuation of such packages to the amount of the overall budget constraint $b_i$. For example, a telecom in a spectrum auction market might have a very high value for the package of all spectrum licenses in the market, but it has only one million dollar budget available. Therefore, we assume that his value, i.e., willingness-to-pay, for the package of all licenses is $v_i=$\$1 million.\footnote{A generalized assignment market which allows bids on individual objects and with overall budget constraints is different. If bidders cannot reveal their willingness-to-pay for all packages, but only the valuations for individual objects, then the utility of a buyer can be described as a 0/1 knapsack problem. We do not consider this setting further in our article, but focus on the more general combinatorial markets, which allow for the submission of XOR package bids on all possible packages.} 
In summary, the value bidders' utility function is $u_i(a)=v_i(a)$ if $p_i(a) \leq v_i(a)$, and $u_i(a)=- \infty$ otherwise. This means, utility is non-transferable between the value bidders in our model. The values $v_i$ are assumed to be monotone non-decreasing (free disposal) and they are normalized with $v_i(\emptyset)=0$, $\forall i \in I$. 

Note that value bidders would not bid beyond their valuation $v_i(a)$, even if their overall budget $b_i$ is not binding. For example, an advertiser on a digital advertising exchange does not want to have an allocation which is within his overall budget $b_i$, but where he has to pay more for every impression than what the net present value for these impressions is. In spectrum auctions, principals typically determine a budget for different packages, which is based on the net present value of the licenses in the package. The management needs to consider these limits and cannot bid beyond.

We consider an offline environment, where we can match bidders to objects in a single step. 
We will start discussing the simplest variant of a combinatorial auction market in our paper: the multi-unit package auction. In a multi-unit auction we have $m$ identical units of an item. We use the notation of an $n \times m$ auction to point to a multi-unit auction with $n$ bidders and $m$ identical items or units. 
In a multi-unit auction we use the notation $(s_1,s_2, \ldots ,s_n)$ to denote an allocation in which $s_i$ units are assigned to bidder $i$. We focus on package auctions, because without package bids, bidders cannot express their preferences for complements or substitutes, which can lead to arbitrarily low revenue with general valuations.
Moreover, in a multi-unit market we use $v_i(s)$ to describe the valuation of bidder $i$ on $s$ units. When an auctioneer presents a bidder with a package $s$, and the bidder responds with his value $v_i(s)$, we will also refer to this as a \textit{value query} \citep{Nisan06}. The optimization goal is to find an allocation of objects to the bidders, where bidder $i$ gets $s_i$ units, with $\sum_i s_i \leq m$, that maximizes revenue $\sum_i v_i(s_i)$. 
Let us now introduce relevant terms from mechanism design.

\begin{definition}
A (direct revelation) \textit{mechanism} is a social choice function $f:V_1 \times ... \times V_n \to A$ and a vector of payment functions $p_1, ...,p_n$, where $p_i: V_1 \times ... \times V_n \to \Re$ is the amount that player $i$ pays.
\end{definition}

$V_i \subseteq \Re^A$ describes the set of possible valuation functions for bidder $i$.
We sometimes refer to the social choice function as the allocation rule of a mechanism. 

\begin{definition}
A mechanism $(f, p_1, ...,p_n)$ is called \textit{incentive compatible} if for every bidder $i$, every $v_1 \in V_1, ...,v_n \in V_n$ and every $v'_i \in V_i$, if we denote $a=f(v_i,v_{-i})$ and $a'=f(v'_i,v_{-i})$, then $u_i(a) \geq u_i(a')$. 
\end{definition}

We will also talk about a \textit{truthful} or \textit{strategy-proof} mechanism in this context when truthtelling is a dominant strategy.

Desirable goals in mechanism design are \textit{Pareto optimality}, the maximization of \textit{social welfare}, and \textit{revenue}. 
Utilitarian social welfare functions add up the value of each individual in order to obtain society's overall welfare. As indicated earlier, the notion of social welfare is difficult if bidders have budget constraints. In this paper we focus on maximizing the auctioneers' revenue. In other words, $v_i(a)$ is the willingness-to-pay of a bidder $i$ for a package $a$ (eventually trimmed to $b_i$), and the auctioneer wants to maximize the sum of these values in the allocation: $f_{R}=\max_{a \in A}\sum_{i \in I_0}{v_i(a)}$. Assuming that every bidder is telling the truth, $f_R$ states the optimal revenue or is optimal, for short.  


The other desirable goal in mechanism design is Pareto-optimality.
\begin{definition}
A pair of allocation and payments $(a, p_1,\ldots,p_n)$ is Pareto-optimal if for no other pair $(a',p_1',\ldots,p_2')$ are all bidders and seller better off, $u_i(a')\geq u_i(a)$, including the seller $\sum_{i \in I} p_i' \geq \sum_{i \in I} p_i$, with at least one of the inequalities strict.
\end{definition}
A quasi-linear mechanism is Pareto-optimal if in equilibrium it selects an allocation or choice $a$ such that $\forall i \forall a'$, $\sum_i v_i(a) \geq \sum_i v_i(a')$. Therefore, an allocation that solves the social welfare maximization problem is Pareto optimal. In a non-quasi-linear environment with value bidders, maximizing social welfare is a sufficient, but not a necessary condition for Pareto-optimal allocations with value bidders. 

\begin{example}\label{swf_vs_efficiency}
To see this consider an example with  bidder 1 interested in item $A$ for \$8 and $B$ for \$5, and a bidder 2 with a value of \$7 for $A$ and \$6 for $B$. If the auctioneer allocates $B$ to bidder 1 and $A$ to bidder 2 with the price equal to their bid, then the utility of the auctioneer and the two bidders would be $(12, 5, 7)$ rather than $(14, 8, 6)$ in the social welfare maximizing allocation. Both allocations are Pareto-optimal, though. 
\end{example}


A mechanism is \textit{individually rational} if bidders always get nonnegative utility \citep{Nisan07}. 
Most of our analysis focuses on individually rational and truthful approximation mechanisms for value bidders. The algorithmic problem of finding the optimal social welfare for general valuations in  combinatorial auctions is $O(\sqrt{m})$ \citep{Halldorsson00}, which is a natural upper bound on the approximation factor of truthful approximation mechanisms. For quasi-linear bidders randomized approximation mechanisms with the same approximation ratio have been found \citep{Lavi11, Dobzinski12random}. 

The algorithmic problem of allocating multiple units of an item to multi-minded bidders reduces to the knapsack problem, for which a simple greedy algorithm proves an approximation ratio of 2 \citep{Lawler79, Lavi11}. Just like for the knapsack problem, the algorithmic allocation problem can be approximated arbitrarily well and has an FPTAS: approximation ratio of $1+\epsilon$ obtained in time that is polynomial in $n$, log $m$, and $\epsilon^{-1}$. For quasi-linear bidders, the framework by \cite{Lavi11} can be used such that any approximation algorithm witnessing an LP integrality gap can be transformed into an algorithm that is truthful in expectation. For the multi-unit auction problem the integrality gap is 2 and, hence, the framework of Lavi and Swamy gives a 2-approximation. \cite{Dobzinski09} presented an FPTAS for multi-unit auctions that is truthful in expectation for quasi-linear bidders. 
\section{Pareto-optimality and revenue maximization} 

We will first analyze if revenue maximization and Pareto efficiency can be implemented in dominant strategies with value bidders. 
Due to the revelation principle, we limit ourselves to direct revelation mechanisms.





\subsection{Truthful and Pareto-optimal mechanisms} 
Because our setting is similar to the setting without money, achieving strategy-proof and Pareto-optimal mechanisms might seem impossible at first sight. However, payments are available and provide an escape route from the many impossibility results in mechanism design without money.
In the following we show that there exists a Pareto-optimal and strategy-proof mechanism for the problem, which is a simple greedy algorithm similar in spirit to \citep{Lehmann02}.

\begin{definition}[PO auction]
Given a set of items $J$ and bidders $I$, find the $i \in I$ and $S \subseteq J$ with the highest $v_i(S)$.  
Allocate $S$ to $i$ at the price equal to $v_i(S)$, and recurse on ($J \setminus S$) and ($I \setminus \{i\}$).
\end{definition}

For the following theorem, we assume that bidders have strict valuations. That is, for all bundles $S$, and $T$ in the bidder $i$'s demand set, we have that $v_i(S) \ne v_i(T)$, for all bidders $i \in I$. This is a reasonable assumption for example in spectrum auctions, where it is unlikely that two different packages have the same value. We will discuss truthful mechanisms for general valuations including ties in Section \ref{det-mech-gen-val}.

\begin{theorem} 
The PO auction is a deterministic, strategy-proof, and Pareto-optimal mechanism for value bidders with strict valuations.  
\end{theorem}




\proof{} 
Let $W$ denote the set of winning bidders who get a bundle  when the auction ends. 
Assume w.l.o.g. that bidder $1$, bidder $2$, \ldots, bidder $|W|$ are added to $W$ (removed from $I$) in order in subsequent iterations. 
We prove that no other allocation Pareto-dominates the allocation chosen by the algorithm.

In any other allocation, what bidder $1$ gets must be the same as what is assigned to him by the algorithm, 
because he has already received his most preferred bundle and his preferences are strict. 
Thus, changing the bundle will decrease his utility. 
As a result, we consider the reallocation of other items not assigned to bidder $1$. 
But these items are exactly equal to $J$ at the second iteration of the algorithm. 
We can repeat now the same argument for bidder $2$ and his received bundle. 
Continuing with this line of arguments for $|W|$ iterations, we can conclude that no other allocation Pareto-dominates the allocation chosen by the algorithm. 
Moreover, because bidders pay their bid, the utility of the auctioneer is maximized under the current allocation. 
Thus, the outcome of the mechanism is Pareto-optimal.

The mechanism is truthful because at each iteration of the algorithm, 
the only way to get a package by a bidder is to bid higher than his true budget in order to have the highest bid for any bundle of available items, which leads to a negative utility in case of winning. 
Bidding lower cannot increase the chances of winning either. 
\endproof

It is easy to see that the mechanism achieves the best possible revenue in multi-unit markets with linear valuations. However, the revenue of the mechanism in general can be as low as $\frac{1}{n}$ of the optimal revenue (see Example \ref{low_revenue}).

\begin{example}\label{low_revenue}

Suppose in a multi-unit market, bidders have valuations $v_i(1)=x+\epsilon, v_i(2)=x+2\epsilon, \ldots, v_i(m)=x+m\cdot \epsilon$, for any $\epsilon>0$, and $i \in I$. With these valuations, the mechanism will return revenue of $x+m\cdot\epsilon$. But the optimal revenue can be higher than $n \cdot x$. Thus, the mechanism returns a result with a revenue lower than $\frac{1}{n}$ of the optimal revenue. 

\end{example}

Note that if the mechanism chooses the revenue-maximizing allocation based on the bids it would not be strategy-proof any more because bidders could shade lower-valued bids in order to get a higher valued package, as we will discuss in the next section. We will discuss the goal of revenue maximizing mechanisms next.

\subsection{Truthful mechanisms maximizing revenue} 

Let's first consider the trivial case of a single item only. A direct pay-as-bid revelation simply gives the item to the highest bidder. Reporting the truth is a weakly dominant strategy for this mechanism. All bidders will report $v_i(a)$, since they pay what they bid but have no value for payoff. The losing bidders cannot gain from decreasing their bid, but would risk making a loss if they bid beyond their valuation. The same result can be extended to \textit{single-minded} and \textit{single-valued} value bidders.

\begin{definition}\citep{Lehmann02}
A valuation $v$ is called single-minded if there exists a bundle of items $S^*$ and a value $v^* \in \mathbb{R}$ such that $v(S)=v^*$ for all $S \supseteq S^*$ and $v(S)=0$ for all other $S$. A single-minded bid is the pair $(S^*, v^*)$.
\end{definition}

\begin{definition} \citep{Babaioff09}
A bidder $i$ is a single-valued (multi-minded) value bidder if there exists a real value $v_i > 0$ such that for any bundle $S \in J$, $v_i(S) \in \{0,v_i\}$ and $v_i(T)=v_i$ for all $T \supseteq S$ if $v_i(S)=v_i$. Both the bidder's value and his collection of desired bundles are assumed to be private information, known only to the bidder himself.
\end{definition}

We will refer to single-minded or single-valued value bidders as \textit{single-parameter} value bidders.

\begin{definition} 
\citep{mu2008truthful} A social choice function with single-parameter value bidders is called monotone in $v_i$, if when valuation $v_i$ makes bidder $i$ win, then so will every valuation $v'_i \ge v_i$.
\end{definition}

\begin{theorem} \label{thm-single-parameter}
A mechanism $(f, p_1, \ldots, p_n)$ with single-parameter value bidders is incentive compatible if and only if the following conditions hold:
\begin{enumerate}
\item $f$ is monotone in every $v_i$.
\item Every winning bidder pays his bid.
\end{enumerate}
\end{theorem}
\proof{} 
(If part) When a bidder has to pay his bid, he has no incentive to bid higher than his true valuation. Also, the bidder has no incentive to bid lower because of two reasons. First, he has no value for the leftover money as he is a value bidder. Second, monotonicity of the mechanism implies that the bidder will never be better off by bidding lower than the true valuation.

(Only-if first part) If $f$ is not monotone then for some $v'_i < v_i$ we have that the bidder wins while he loses with $v_i$. Then the mechanism will not be truthful since the bidder is incentivized to lower his bid from $v_i$ to $v'_i$.

(Only-if second part) Assume a bidder pays $p$ which is less than a winning bid $v'_i$, i.e. $p < v'_i$. This $v'_i$ might be an untruthful bid. The mechanism cannot verify it. The true valuation might have been losing $v_i$, where $v_i=p<v'_i$. Thus, without paying the bid, the bidder is incentivized to bid higher to ensure winning, harmlessly.
\endproof


Notice that with single-minded or single-valued value bidders the winner determination problem to select the revenue-maximizing allocation is still $NP$-hard \citep{CaBookLehmann06}. Theorem \ref{thm-single-parameter} allows us to use the existing monotone approximation algorithms for the allocation problem in markets with single-minded or single-valued value bidders which are such that the computational hardness cannot be ignored (for such approximation algorithms, refer to \cite{Nisan07}). With a pay-as-bid payment rule such single-parameter value bidders would not have an incentive to deviate from truthful bidding. Unfortunately, these positive results do not carry over to multi-minded value bidders with general valuations. 

\begin{theorem} \label{thm-impossibility}
There is no strategy-proof and revenue maximizing auction mechanism for general value bidders and more than one object for sale.
\end{theorem}
\proof{} 
Suppose there is a supply of 2 homogeneous units of an item. Bidder 1 has a value $v_1(1)=x$ and $v_1(2)=x+\epsilon$, $\epsilon>0$, while bidder 2 has $v_2(1)=x$. 
The allocation maximizing revenue $f_R$ is to assign one unit to bidder 1 at price $x$ and one unit to bidder 2 at the similar price with an optimal revenue of $2x$. Bidder 1 can increase his utility to $x+\epsilon$, by bidding on two units only and pretending that his value for 1 unit is null. This way $f_R$ allocates two units to bidder 1. Thus, bidder 1 has an incentive to shade his bid for one item.
Payment rules other than a pay-as-bid rule do not set incentives for value bidders to bid truthfully because value bidders do not maximize payoff but value. 
\endproof

In the quasi-linear setting, assignment markets allow even for strategy-proof ascending auctions \citep{Demange86}. Unfortunately, the negative result in Theorem \ref{thm-impossibility} even holds for assignment markets, where value bidders can only win at most one from multiple items. 

\begin{corollary} \label{thm-assignment}
There is no strategy-proof and revenue maximizing auction mechanism for value bidders in assignment markets.
\end{corollary}
\proof{} 
Suppose there is a market with two items $A$ and $B$ and two bidders. Bidder 1 has a value of $x$ for item $B$, while bidder 2 has a value of $x$ for item $A$ and a value of $x+\epsilon$ for item B. Bidder 2 can increase his utility by bidding $0$ for item $A$, which would make him win item $B$ and lead to a revenue of $\frac{1}{2}$ of the optimal.
\endproof

One escape route from these negative results on strategy-proof and revenue maximizing auctions is to give up on optimal solutions and restrict attention to approximation mechanisms to achieve strategy-proofness. In other words, we try to keep strategy-proofness at the expense of optimal revenue. There is a growing literature on approximation mechanisms for quasi-linear bidders \citep{Lavi07} and it is a natural question to understand approximation mechanisms for value bidders. 


\section{Approximation mechanisms for markets with two bidders} \label{approx_for_two_bidders}
In this section, we deal with truthful approximation mechanisms for multi-minded bidders. We impose truthfulness as a constraint and analyze if mechanisms with a good approximation ratio of the optimal revenue are possible. 
We will first focus on $2 \times 2$ markets and introduce an assumption when good approximation ratios are possible. We will refer to this as the public strongest bidder assumption which is defined as follows.

\begin{definition}[Public strongest bidder - PSB]
With public strongest bidder (or for short, PSB) assumption, it is publicly known which bidder has the highest bid on all items. This means, when bidder $k$ is the public strongest bidder it is publicly known that $v_k(J) \geq v_i(J)$, $k \neq i \in I$.
\end{definition}

While this might look like an strong assumption at first, such asymmetries in a market can often be observed \citep{Maskin00}. For example, in many spectrum auction markets, there is a national carrier and it is public knowledge that he will have the highest willingness-to-pay for his most desired package \citep{Klemperer02}. In the following, we assume that bidder 1 is the strongest bidder, when the PSB assumption is satisfied.

First, we show that there is a strategy-proof polynomial-time mechanism with a good approximation ratio for the $2 \times 2$ market ($2$ bidders and $2$ units) assuming the PSB assumption. The golden ratio mechanism for two units of a homogeneous item and two bidders works as follows.

\begin{definition}[Golden ratio mechanism]
\begin{enumerate}

\item Case 1: $v_1(2) > v_2(2)$. If $v_2(1) > \frac{\sqrt 5 - 1}{2}v_1(2)$ then assign one item to each bidder. Otherwise, assign both items to bidder 1. The winners pay what they bid for the units assigned to them.

\item Case 2: $v_1(2) = v_2(2)$. If there is any $i \in \{1,2\}$ such that $v_i(1) > \frac{\sqrt 5 - 1}{2}v_i(2)$, then assign one item to each bidder. Otherwise, assign both items to one of the bidders, uniformly at random.
\end{enumerate}
\end{definition}

Note that the approximation ratio $r=\frac{\sqrt 5 - 1}{2}$ can be derived from the golden ratio, which is $1/r$. Note that we always assume free disposal, such that two items are worth at least the value of one item. 

The PSB assumption is required for the golden-ratio mechanism. For example, for the true valuations shown in Table \ref{imp_tab_golden_ratio}, the golden ratio mechanism assigns both units to bidder 1.

\begin{table}[!htbp] 
\centering
\begin{tabular} {c|c|c} 
   & 1 & 2 \\
\hline
1 &  $64$ & \cellcolor[gray]{0.7} $100$\\
\hline
2 & $55$ & $56$\\

\end{tabular}
\caption{} \label{imp_tab_golden_ratio}
\end{table}

But without the PSB assumption bidder 2 could get one unit by pretending $v_2(2)=101$, because the mechanism will assign one unit to each bidder.


\begin{theorem} The golden ratio mechanism for the $2 \times 2$ market is strategy-proof and its revenue is at least $\frac{\sqrt 5 - 1}{2}$ of the optimal revenue under the PSB assumption.

\end{theorem}
\proof{} 
The mechanism executes either Case 1 or Case 2. Due to the PSB assumption, the bidders cannot decide which case to be executed. Thus, we can consider the revenue ratio and truthfulness of each case, separately.

Let $r = \frac{\sqrt 5 - 1}{2}$. It is easy to see that in every assignment by Case 1 of the mechanism the lower bound of the optimality is respected. 

In terms of truthfulness, in Case 1 if any bidder wants to get a more valuable package, he has to increase the bid for that package which means he has to pay more than his true valuation. Thus, no bidder has an incentive to change the bids. Without the PSB assumption, bidder 2 could pretend to be the high bidder on two units with $v_1(1) > \frac{\sqrt 5 - 1}{2}v_2(2)$, in order to win one unit rather than nothing at all.

In the tie-breaking situation handled in Case 2 of the auction, the assignment of one unit to each bidder obviously respects the ratio. In case any of the bidders gets both units, still the ratio is respected. Assume w.l.o.g. bidder $1$ gets both units. For every $i \in \{1,2\}$, we have that $v_i(1) \leq r\cdot v_1(2)$, thus $\sum_{i \in \{1,2\}} v_i(1) \leq  2r\cdot v_1(2)$, then $v_1(2) \geq \frac{1}{2r}\sum_{i \in \{1,2\}} v_i(1) \geq r\cdot \sum_{i \in \{1,2\}} v_i(1)$. 

Case 2 of the mechanism is also truthful. Suppose both bidders get one item because for bidder 1 we have $v_1(1) > r\cdot v_1(2)$. Now if he wants to get both items, he has to bid lower for one item. But, what he gets then is $\frac{1}{2} v_1(2)$ in expectation, and $\frac{1}{2} v_1(2) < r\cdot v_1(2)<v_1(1)$. Moreover, when any of the bidders gets two units randomly, if a bidder wants to get one unit for sure, he has to bid beyond his true valuation on one unit which results in a loss in case of winning. Thus, in case of a tie, bidders have no incentive to lie about their valuations as well. 

\endproof

As the next theorem shows, this ratio of optimal revenue is tight and no deterministic, and truthful mechanism can achieve more.

\begin{theorem}
There is no deterministic mechanism for the $2 \times 2$ market which is strategy-proof and its revenue is more than $(\frac{\sqrt 5 - 1}{2}+ \epsilon)$ optimal even under the PSB assumption for any $\epsilon > 0$.
\end{theorem}
\proof{} 

We set $r = \frac{\sqrt 5 - 1}{2}$. Consider the valuations shown in Table \ref{imp_tab_2items_1}.\footnote{Throughout the paper when valuations are shown in tables, the rows are representing bidders and columns are the number of units and thus each cell shows the valuation of the corresponding row bidder for the number of units in the column. For example, in the Table \ref{imp_tab_2items_1} the valuation of bidder 2 for one unit is $v_2(1)=r\cdot x+2\epsilon$.}

\begin{table}
\begin{minipage}[b]{.50\textwidth}
  \centering
		\begin{tabular} {c|c|c} 
		   & 1 & 2 \\
		\hline
		Bidder 1 & \cellcolor[gray]{0.7} $x-\epsilon$ & $x$\\
		\hline
		Bidder 2 & \cellcolor[gray]{0.7} $r\cdot x+2\epsilon$ & $r\cdot x+ 3\epsilon$\\
		\end{tabular}
  \caption{}  \label{imp_tab_2items_1}
\end{minipage}\qquad
\begin{minipage}[b]{.50\textwidth}
  \centering
		\begin{tabular} {c|c|c}
		   & 1 & 2 \\
		\hline
		Bidder 1 & \cellcolor[gray]{0.7} $0$ & $x$\\
		\hline
		Bidder 2 & \cellcolor[gray]{0.7} $r\cdot x+2\epsilon$ & $r\cdot x+ 3\epsilon$\\
		\end{tabular}
  \caption{}   \label{imp_tab_2items_2}
\end{minipage}
\end{table}



In this case, giving both items to bidder 1 will result in a less-than-$r$ ratio of the optimal revenue, since

\begin{displaymath}
\frac{x}{r\cdot x+2\epsilon+x-\epsilon} < \frac{x}{r\cdot x+x} = r.
\end{displaymath}

Thus, we choose to give one item to each bidder. Now consider the new valuations shown in Table \ref{imp_tab_2items_2}.



Here, the best revenue is presented by the assignment of two items to bidder 1, but this cannot happen because then bidder 1 has an incentive to pretend having valuations as in Table \ref{imp_tab_2items_2} when his true valuations are as given in Table \ref{imp_tab_2items_1}. Thus, the best possible ratio of revenue in this case will be $\frac{r\cdot x+2\epsilon}{x} \simeq r+\epsilon = \frac{\sqrt 5 - 1}{2}+\epsilon$. 

\endproof

Randomization is often used in the design of quasi-linear mechanisms. In what follows, we  present a randomized truthful mechanism for $2 \times 2$ markets which yields a better approximation factor. As usual, we say a mechanism is \textit{truthful in expectation}  if for every bidder $i$, every $v_1 \in V_1, ...,v_n \in V_n$ and every $v'_i \in V_i$, if we denote $a=f(v_i,v_{-i})$ and $a'=f(v'_i,v_{-i})$, then $E[u_i(a)] \geq E[u_i(a')]$.

\begin{definition}[Randomized $2\times 2$ Mechanism] \label{rand-mech-2-2}
\begin{enumerate}
\item If $v_i(2) > v_j(2)$ with $i,j \in I$, then set $q = \frac{v_j(1)}{v_i(2)}$. Assign each bidder a single unit with probability $q$ and assign both units to bidder $i$ with probability $1-q$. The winners pay what they bid for the units assigned to them.
\end{enumerate}
\end{definition}

\begin{theorem} The randomized $2\times 2$ mechanism is truthful in expectation and its expected revenue is at least $\frac{3}{4}$ optimal revenue.
\end{theorem}

\proof{}

First, we discuss the revenue ratio. We assume w.l.o.g that $v_1(2)>v_2(2)$. Let us use a simpler notation by setting $\beta=v_1(1)$, $\theta=v_1(2)$, and $\alpha=v_2(1)$. 

\begin{table}[!htbp] 
\centering
\begin{tabular} {c|c|c}
   & 1 & 2 \\
\hline
Bidder 1 & $\beta$ & $\theta$\\
\hline
Bidder 2 & $\alpha$ & \\

\end{tabular}
\caption{}\label{rand_tab_2items_2}
\end{table}

In order to see the ratio of the mechanism we consider two cases: $\theta \geq \alpha + \beta$, and $\theta < \alpha + \beta$.  First, $\theta \geq \alpha + \beta$. 
Here the ratio of the mechanism will be

\begin {displaymath}
\frac{\frac{\alpha}{\theta}(\alpha + \beta)+(1-\frac{\alpha}{\theta})\theta}{\theta}  \geq \frac{\alpha^2}{\theta^2}-\frac{\alpha}{\theta}+1\geq \frac{3}{4}.
\end {displaymath}

The first inequality is because $\beta \geq 0$. Second, $\alpha + \beta > \theta$. Then the ratio of the mechanism will be 
\begin {displaymath}
\frac{\frac{\alpha}{\theta}(\alpha + \beta)+(1-\frac{\alpha}{\theta})\theta}{\alpha+\beta} \geq \frac{\alpha}{\theta}+\frac{\theta-\alpha}{\theta+\alpha} \geq 2(\sqrt{2} - 1).
\end {displaymath}

The first inequality is because $\beta \leq \theta$, and the second inequality is because the minimum of the expression occurs when $\alpha = (\sqrt{2}-1)\theta$. Thus, the minimum ratio will be $\frac{3}{4}$.

The mechanism is truthful as it is shown in the following. The bidder who has the higher valuation for two units will get one unit with probability one.
If a bidder wants to increase his bid for two units beyond his value to become the highest bidder for two items, he will also incur a loss with positive probability. If the strong bidder on two units submits a bid that is different from his value for one unit, this does not impact the probability of winning two units. 

If the weak bidder on two units submits a bid on one unit, which is higher than his value he might incur a loss with positive probability. If this bidder bids less than his value, he just decreases his chances of winning even a single unit. 

\endproof

Notice, that for the randomized mechanism, no PSB assumption was necessary. 

Having the golden ratio mechanism under the PSB assumption, there is hope to find deterministic approximation mechanisms for markets with more items or more bidders as well. But, unfortunately as the following theorem states the mechanism is not extensible even for $2 \times 3$ markets.

\begin{theorem}
There is no mechanism for the $2 \times 3$ market which is deterministic, strategy-proof, and its worst-case revenue is more than $(\frac{1}{2}+ \epsilon)$ optimal even under the PSB assumption.
\end{theorem}

\proof{} 

Suppose there is. Consider the following valuations in Table \ref{imp_tab_3items_1} with two bidders and three homogenous items.

\begin{table}
\begin{minipage}[b]{.50\textwidth}
  \centering
\begin{tabular} {c|c|c|c} 
   & 1 & 2 & 3 \\
\hline
Bidder 1 & $x$ &  \cellcolor[gray]{0.7} $x+ \epsilon$ & $x+ 3\epsilon$\\
\hline
Bidder 2 & \cellcolor[gray]{0.7} $x$ & $x+ \epsilon$ & $x+ 2\epsilon$\\

\end{tabular}
  \caption{}  \label{imp_tab_3items_1}
\end{minipage}\qquad
\begin{minipage}[b]{.50\textwidth}
  \centering
\begin{tabular} {c|c|c|c}
   & 1 & 2 & 3 \\
\hline
Bidder 1 & $x$ &  $x+ \epsilon$ &  \cellcolor[gray]{0.7} $x+ 3\epsilon$\\
\hline
Bidder 2 & $0$ & $x+ \epsilon$ & $x+ 2\epsilon$\\

\end{tabular}  \caption{}  \label{imp_tab_3items_2}
\end{minipage}
\end{table}



The mechanism must assign one unit to one bidder and two units to the other bidder, otherwise the mechanism returns a low ratio of $\frac{1}{2}+\epsilon$ and the theorem holds. Assume w.l.o.g one unit is assigned to bidder 2 and two units to bidder 1. Now, consider the new valuations in Table \ref{imp_tab_3items_2}.



If bidder 2 obtains two or three units by bidding 0 for one item, he would have an incentive to lie when his true valuations were as given in Table \ref{imp_tab_3items_1}. 
Therefore, the only solution of the new valuations is the assignment of three units to bidder 1. However, this valuation is only $(\frac{1}{2}+ \epsilon)$ optimal. If bidder 2 is still assigned 1 unit and bidder 1 is assigned two units as for Table \ref{imp_tab_3items_1}, the revenue is also only $(\frac{1}{2}+ \epsilon)$ optimal, and the auctioneer could be made better off, by allocating three units to bidder 1, without decreasing the utility of any other participant. Therefore, there is no truthful mechanism, which can achieve a higher approximation factor. 

\endproof

With a symmetric argument we get the following result for a market where bidders are interested in two items and the bundle of both items.

\begin{corollary}
There is no mechanism for two bidders and two heterogeneous items which is deterministic, strategy-proof, and with a worst-case revenue more than $(\frac{1}{2}+ \epsilon)$ optimal even under the PSB assumption.
\end{corollary}

Consequently, the best possible approximation factor achievable by a deterministic mechanism for two bidders and two items is $(\frac{1}{2}+ \epsilon)$ which is what we can gain by the assignment of all items to the strongest bidder. 

\section{Deterministic mechanisms for general value bidders} \label{det-mech-gen-val}

In this section, we characterize properties of truthful revenue maximizing mechanisms for value bidders in general. This allows us to answer the question whether an approximation ratio better than $\frac{1}{n}$ is possible for markets with $n$ bidders. This is what we can achieve when bidders are only allowed to submit bids on the grand bundle (the bundle of all items). We try to analyze this by characterizing the allocation rule of all possible truthful mechanisms. We present our negative results for multi-unit auctions which are a subset of combinatorial auctions.

Generally speaking, with value bidders at most one value query for each bidder can be verified by a truthful mechanism. This severely restricts the possibility of designing deterministic truthful mechanisms. One observation from the golden ratio mechanism and the randomized $2 \times 2$ mechanism is that only a single value query for one package $S$ is used by the auctioneer, i.e., only a single value $v_i(S)$ of each bidder is considered and then the bidder can either win this package or a lower valued one. 
An example for another allocation rule $f$ in a $3 \times 4$ market which satisfies the same properties is: ''if $v_1(4) > v_2(2)+v_3(2)$ then $(4,0,0)$ else $(0,2,2)$.'' In this example, each bidder can only win the package, which is evaluated in the condition of the allocation rule or the empty set. No bidder has an incentive to lie in the value query, because this package is assigned to the bidder. 

\subsection{Allocation rule revisited}
We restrict our attention to those valuations of bidders which make the allocation rule assign all units to only one bidder. These valuations play an essential role in defining the outcome of the allocation rule and we discuss this further in this section. We first define two new notations. Let $v=(v_1, \ldots, v_n)\in V_1 \times \ldots \times V_n$ point to an arbitrary set of valuations. We denote by $(i \hookleftarrow s)$ the assignment in which $s$ units are assigned to bidder $i$. Given an allocation rule $f:V_1\times \ldots \times V_n \to A$, we define a new function  
\begin{displaymath}
  \begin{array}{l l}
F_i:V_1 \times \ldots \times V_n \to \{\text{\emph{true, false}}\} & \\
 F_i(v) = \left\{ 
  \begin{array}{l l}
    \text{\emph{true}} & \quad \text{if $f(v)=(i \hookleftarrow m)$,}\\
    \text{\emph{false}} & \quad \text{otherwise}.
  \end{array} \right. &
  \end{array}
\end{displaymath}

Intuitively speaking, $F_i$ determines whether a set of bidders' valuations will result in the assignment of $m$ units to bidder $i$. Notice, $F_i(\cdot)$'s are disjunctive, i.e. $F_i(v)$ $\&$ $F_j(v) = \text{\emph{false}}$, $\forall i,j \in I$, because the grand bundle cannot be assigned to more than one bidder, simultaneously.
Now, using $F_i$'s we redefine the allocation rule as follows. 
\begin{displaymath}
  \begin{array}{l l}
f^*: V_1\times \ldots \times V_n \to A & \\
 f^*(v) = \left\{ 
  \begin{array}{l l}
     (i \hookleftarrow m) & \quad \text{if $F_i(v) = \text{\emph{true}}$, $\forall i \in I$},\\
    f(v) & \quad \text{otherwise}.
  \end{array} \right. &
  \end{array}
\end{displaymath}

\begin{example} \label{example_F_i}
Consider the following allocation rule for the $3\times 4$ market.
\begin{enumerate}
\item If $v_1(4) > \max(v_2(4),v_3(4))$ $\wedge$ $(v_1(4) > v_2(2)+v_3(2))$ then $(4,0,0)$. 
\item If $v_2(4) > \max(v_1(4),v_3(4))$ $\wedge$ $(v_2(4) > v_1(2)+v_3(2))$ then $(0,4,0)$.
\item If $v_3(4) > \max(v_1(4),v_2(4))$ 
\item If $v_1(4) > \max(v_2(4),v_3(4))$ then $(0,2,2)$.
\item If $v_2(4) > \max(v_1(4),v_3(4))$ then $(2,0,2)$.
\end{enumerate}

In this allocation rule, the functions $F_i(\cdot)$'s are defined as follows. 
$F_1(v) \equiv$ $v_1(4) > \max(v_2(4),v_3(4))$ $\wedge$ $(v_1(4) > v_2(2)+v_3(2))$, 
$F_2(v) \equiv$ $v_2(4) > \max(v_1(4),v_3(4))$ $\wedge$ $(v_2(4) > v_1(2)+v_3(2))$, and 
$F_3(v) \equiv$ $v_3(4) > \max(v_1(4),v_2(4))$. These functions are disjunctive and all might become false simultaneously.
\end{example}

This new rewriting of the allocation rule will be useful for deriving  a general result in mechanism design for value bidders as will be shown in the following.

It is easy to observe that the two definitions of the allocation rule are equivalent. However, for the sake of completeness, we give the proof in the following.

\begin{lemma} \label{lem-two-rules-equal}
For any set of valuations $v \in V_1 \times \ldots \times V_n$, we have that $f^*(v)=f(v)$.
\end{lemma}
\proof{}
Consider an arbitrary set of valuations $v$. To show the claim, we must prove that $f^*(v)=f(v)$. We divide the argument into two cases. 
\begin{enumerate}

\item $\exists i \in I$ to whom all units are assigned. By definition, $f(v) = (i \hookleftarrow m)$. Then, $F_i(v)=\text{\emph{true}}$, per definition. Thus, $f^*(v)=(i \hookleftarrow m)$, the desired conclusion.

\item $\nexists i \in I$ to whom all units are assigned. Then, $F_i(v)=\text{\emph{false}}$, $\forall i \in I$, per definition. Then again per definition, $f^*(v)=f(v)$, the desired conclusion. 

\end{enumerate}
\endproof

That $f^*$ and $f$ are equivalent, lets us focus on $f^*$ and try to find conditions which $f^*$ must satisfy in order to achieve a good revenue as well as obtaining truthfulness.

\subsection{Properties of $F_i(\cdot)$}
The domain of function $F_i(\cdot)$ is the set of all valuations. Thus, one might guess that in computing $F_i(\cdot)$, the valuations of bidders for any bundle $j \le m$ might be queried. Yet, in what follows we present multiple lemmata which show that this is not the case and restrict the arguments of $F_i(\cdot)$ to only valuations of bidders for $m$ units: $v_j(m)$ $\forall j \in I$.

We first look at the lemma which describes properties of mechanisms, which avoid low revenue.

\begin{lemma} \label{lem-all-valuations-on-grand-bundle}
 In order to avoid arbitrarily low revenues of less than $\frac{1}{n}$, arguments of function $F_i(\cdot)$ must include all value queries $v_j(m)$, $\forall j \in I$, i.e. $F_i(\cdot)$ has to be a function of all $v_j(m)$'s. 
\end{lemma}
\proof{}
We show that, in case, there exists a bidder whose valuation for $m$ units is not queried in computing $F_i(\cdot)$ then $f^*$ might end up in an allocation with arbitrarily low revenue of less than $\frac{1}{n}$. We divide the argument into two cases.
\begin{enumerate}
\item $v_i(m)$ is not queried in $F_i(\cdot)$. Thus, the mechanism decides if all items are assigned to bidder
$i$ without evaluating his valuation. Now consider a situation in which $F_i(\cdot)=\text{\emph{false}}$. Here, bidder $i$ does not get the grand bundle $m$. However, he can have an arbitrarily high valuation for getting the bundle. Thus, the revenue of the mechanism can become arbitrarily low in comparison to the optimal one.

\item $v_j(m)$, $j \ne i$ is not queried in $F_i(\cdot)$. This $v_j(m)$ can be arbitrarily high. Thus, without considering it, in case of $F_i(\cdot)=\text{\emph{true}} $, the revenue-ratio can become arbitrarily low.

\end{enumerate}

	Therefore, $F_i(\cdot)$ has to query all $v_j(m)$'s, $j=1,\ldots,m$. This completes the proof.

\endproof

The second lemma holds only for value bidders. This lemma takes into account the truthfulness of mechanism.
\begin{lemma} \label{lem-one-value-query}
 In order to obtain truthfulness, no valuation from bidder $i$ other than $v_i(m)$ can be queried in computing $F_i(\cdot)$.
\end{lemma}
\proof{}
We prove it by contradiction. Assume apart from $v_i(m)$ there is another valuation of bidder $i$ queried in computing $F_i(\cdot)$. Then, the bidder can manipulate this additional value query to make $F_i(\cdot)=\text{\emph{true}} $. To see how this can happen, let us call this additional variable as $v_i(k)$, $k < m$. Assume w.l.o.g. that $F_i(v_i(m), v_i(k), \ldots) = \text{\emph{false}}$, and $F_i(v_i(m), v'_i(k), \ldots) = \text{\emph{true}} $. Then bidder $i$ will be better off by reporting $v'_i(k)$ rather than $v_i(k)$ and the mechanism cannot set incentives for truthful bidding. 
\endproof

\begin{lemma} \label{lem-equal-but-grand}
If two bidders are equal in all valuations except for the grand bundle, then in case the mechanism wishes to assign the grand bundle to one of them, it must assign the grand bundle to the stronger bidder, otherwise the mechanism will be neither revenue maximizing nor truthful.
\end{lemma}
\proof{}
Assume two bidders $i, j \in I$ are equivalent in everything except for the grand bundle for which $v_i(m) > v_j(m)$. Also, assume that the mechanism assigns the grand bundle to bidder $j$. 

The mechanism is also not revenue maximizing because it can get a higher revenue by assigning the grand bundle to bidder $i$. 

The mechanism is not truthful because the true bid by bidder $j$ might have been $v'_j(m)>v_i(m)$. Thus, under this selection, the stronger bidder is incentivized to shade his bid for the grand bundle.
\endproof

The next lemma proves that, by taking into consideration the truthfulness, if all units are assigned to a bidder, that bidder has to be the strongest on the grand bundle.
\begin{lemma} \label{lem-strongest_bidder}
If $F_i(v)=\text{true}$ then $v_i(m) > \max _{j \ne i} v_j(m)$, $\forall i,j \in I $, otherwise the mechanism is not truthful.
\end{lemma}
\proof{}
We prove it by contradiction. Let $v$ be an arbitrary set of valuations. Let $F_j(v)=\text{\emph{true}} $,  for a bidder $j$. Assume, there exists a bidder $i$ such that $v_i(m) > v_j(m)$. By our assumption $F_i(v) = \text{\emph{false}}$. We must prove that such a mechanism cannot be truthful. 

If the valuations of bidder $j$ were such that $\forall s < m$, $v_j(s) = v_i(s)$,  he certainly had no chance to get the grand bundle because based on Lemma \ref{lem-equal-but-grand} when the mechanism wishes to assign the grand bundle to one of these two bidders, the winner will be bidder $i$. Thus, there could have been a lie by reporting $v'_j(s) \neq  v_i(s)$ for some $s < m$ to get to this status in which $F_i(v) = \text{\emph{false}} $, and $F_j(v)=\text{\emph{true}} $. Now, because all $m$ units are assigned to bidder $j$, the mechanism has no way to prevent his possible lie on smaller number of units, i.e. $v_j(s)$, $\forall{s<m}$. Thus, there exists an incentive for bidders to misreport their valuations on $s$ and therefore the mechanism cannot be truthful.
\endproof

Considering Lemma \ref{lem-strongest_bidder}, we can be more specific about the function $F_i(\cdot)$ as the following lemma states.

\begin{lemma} \label{lem-new-Fi}
The rewriting of function $F_i(\cdot)$ as the following is without loss of generality. $F_i(v) = (v_i(m) > \max_{j \ne i} v_j(m))$ $\wedge$ $F'_i(v)$, where
\begin{displaymath}
  \begin{array}{l l}
F'_i:V_1 \times \ldots \times V_n \to \{\text{true, false}\} & \\
 F'_i(v) = \left\{ 
  \begin{array}{l l}
    \text{true}  & \quad \text{\emph{if} $F_i(v)=\text{true}$} ,\\
    \text{false}  & \quad \text{\emph{if} $F_i(v)=\text{false}$ $\wedge$ $v_i(m) > \max_{j \ne i} v_j(m)$},\\
    \text{\small{don't care}} & \quad \text{$v_i(m) < \max_{j \ne i} v_j(m)$}.
  \end{array} \right. &
  \end{array}
\end{displaymath}
\end{lemma}
\proof{}
Based on Lemma \ref{lem-all-valuations-on-grand-bundle}, in $F_i(\cdot)$ all valuations for the grand bundle must be queried. Thus, the inclusion of all $v_j(m)$'s in the redefinition of $F_i(\cdot)$ follows this very point. We can iterate over the four cases defined by $F_i(v)=\text{\emph{true}} | \text{\emph{false}}$ and $v_i(m) > | < \max_{j \ne i} v_j(m)$. The only case not supported by the new definition is $F_i(v)=\text{\emph{true}} $ and $v_i(m) < \max_{j \ne i} v_j(m)$. But, based on Lemma \ref{lem-strongest_bidder}, this case is impossible. Thus, the new rewriting is without loss of generality, the desired conclusion.
\endproof	

In the next lemma, we show that when a bidder has the highest valuation for the grand bundle, he will get the all units. In order to show this, we draw on a  property of social choice functions, namely anonymity. Anonymity requires that the outcome of a social choice function is unaffected when agents are renamed. 

\begin{lemma} \label{lem-biggest-everything}
If $v_i(m) > \max_{j \ne i} v_j(m)$ then $F_i(v)=\text{true}$, $\forall i,j \in I $, otherwise the mechanism is not truthful.
\end{lemma}
\proof{}
Suppose $\exists i \in I$, and $v \in V_1 \times \ldots \times V_n$ such that $v_i(m) > \max_{j \ne i} v_j(m)$ and $F_i(v)=\text{\emph{false}}$. We prove that such a mechanism will be manipulable. By our assumptions, no bidder gets the grand bundle. Thus, a bidder $j\ne i$ gets $r<m$ units. 

$v_j(r)$ must have been queried in $F_i(v)$ and therefore in $F'_i(v)$. Querying any other valuation $v_j(t)$, $0<t\le m$, $t \ne r$ in $F'_i(v)$, makes the mechanism manipulable. That is bidder $j$ can manipulate the mechanism in the following way. Assume $F'_i(v_j(t),\ldots)=\text{\emph{true}}$ then bidder $j$ can report $v'_j(t)$ such that $F'_i(v'_j(t),\ldots)=\text{\emph{false}}$ and then ending up in the assignment which assigns $r$ units to him. The mechanism has no way to prevent such a lie and therefore it is manipulable. We emphasize that even querying $v_j(m)$ in $F'_i(v)$ makes the mechanism manipulable as follows. If $v_j(m)$ has caused $F'_i(v)=\text{\emph{false}}$, a lower value of $v'_j(m)$ could let it be true. Thus, when the true valuation of bidder $j$ is $v'_j(m)$ he can be better off by reporting $v_j(m)$, instead.
Notice that we have assumed some valuations of any bidder $j$ is queried in $F'_i(v)$ because without this assumption, $F'_i(\cdot)$ is a function without arguments and therefore we have $F_i(v)=\text{\emph{true}}$ when $v_i(m) > \max_{j \ne i} v_j(m)$ and we are done. 

Thus, we learn that when $r$ units are assigned to bidder $j$, no valuation of bidder $j$ rather than $v_j(r)$ can be queried in $F'_i(\cdot)$. Now, consider this from the reverse angle. That one valuation of bidder $j$, i.e. $v_j(r)$, which is queried in $F'_i(\cdot)$, in case $F'_i(v)=\text{\emph{false}}$, becomes winning and the corresponding number of units $(r)$ gets assigned to the bidder. Therefore, for all valuations which do not end up in the assignment of the grand bundle to bidder $i$, the mechanism always assigns a fixed number of units to bidder $j$.

The mechanism is assumed to be anonymous (not depending on the labels of the bidders), therefore for every bidder we get a fixed assignment. Moreover, in an anonymous mechanism bidders are symmetric, therefore every bidder will get a fixed assignment of $\frac{m}{n}$ units. 

Up until now, we learn that the mechanism will either assign the grand bundle to a bidder or will assign $\frac{m}{n}$ units to any bidder
\footnote{This mechanism is already of approximation $\frac{1}{n}$. Let $s=\frac{m}{n}$. Consider a set of valuations where $\forall i \in I$, $v_i(t)=0$, when $t \le s$ and $v_i(t)\cong v_i(m)$, for $t>s$, and all the valuations for the grand bundle are approximately the same. Then if the mechanism assigns either the grand bundle to one bidder or $\frac{m}{n}$ units to each bidder, the approximation ratio will be of $\frac{1}{n}$.}.
Moreover, we know the following about the arguments of any function $F'_i(\cdot)$. The only valuation of bidder $i$ queried in $F'_i(\cdot)$ is $v_i(m)$, based on Lemma \ref{lem-one-value-query}, and the only valuation of any bidder $j$ queried in $F'_i(\cdot)$ is $v_j(\frac{m}{n})$, based on the argument above.

But, such a mechanism is not truthful. Let $s=\frac{m}{n}$. Consider a valuation $v$ where $v_i(m)>v_j(m)>\max_{k \in I\setminus\{i,j\}}v_k(m)$, $F'_i(v)=\text{\emph{false}}$ and $F'_j(v)=\text{\emph{true}}$. This can happen when $v_i(s)=0$, and $v_j(s)\cong v_j(m)\cong v_i(m)$. Notice, it is assumed that $v_i(s)$ is influential in $F'_j(v)$ and thus sometimes can change the value of it. Now, consider a decrement in the value of $v_i(m)$ to $v'_i(m)<v_j(m)$. With this new valuation $v'$, we have $F_j(v')=true$. Notice, $F'_j(v')$ will remain true since it's arguments are untouched. Thus, $v'$ is manipulable by bidder $i$ by increasing $v'_i(m)$ to $v_i(m)>v_j(m)$ in order to get $s$ units. Thus, allowing the assignment of $s$ units to any bidder makes the mechanism manipulable.

This impossibility arose because while $v_i(m) > \max_{j \ne i} v_j(m)$, we assumed $F_i(v)=\text{\emph{false}}$. Thus, if $v_i(m)$ is the highest bid on the grand bundle, we will always have $F_i(v)=\text{\emph{true}}$, the desired conclusion. 
\endproof

\begin{example} \label{example-gen-mech}
Consider the allocation rule given in Example \ref{example_F_i}. Obviously, the allocation rule is not anonymous since the outcome depends on the label of the bidders. If bidder $3$ has the highest valuation for the grand bundle, he will get it but for bidder $1$ and $2$ this does not hold.

In addition, the mechanism is not truthful. Consider a case in which $v_1(4)>v_2(4)>v_3(4)$, $v_1(4)<v_2(2)+v_3(2)$, and $v_2(4)<v_1(4)+v_3(2)$. The outcome of the mechanism in this case will be $(0,2,2)$. But, bidder $1$ can be better off by bidding $v'_1(4)<v_2(4)$, for which the outcome will be $(2,0,2)$.
\end{example}

\begin{theorem} \label{main-val-bidders-imp}
The best revenue ratio achievable by a deterministic and truthful mechanism with value bidders in an $n \times m$ market is $\frac{1}{n}$.
\end{theorem}
\proof{}
According to Lemma \ref{lem-biggest-everything}, the allocation rule assigns everything to the bidder with the highest bid for the grand bundle. This assignment has a worst-case approximation ratio $\frac{1}{n}$.
\endproof

Theorem \ref{main-val-bidders-imp} is easily extensible to combinatorial markets as the following corollary states. Overall, there is a gap between the best approximation ratio of an approximation algorithm to the set packing problem ($O(\sqrt{m})$, and the best truthful approximation mechanism.

\begin{corollary} \label{comb-main-val-bidders-imp}
The best revenue ratio achievable by a deterministic and truthful mechanism with value bidders in a combinatorial market is $\frac{1}{n}$.
\end{corollary}

\section{A randomized mechanism for general value bidders}
\label{general-randomized}
In the previous section, we focused on deterministic mechanisms for value bidders and found an approximation ratio of only $\frac{1}{n}$. It is interesting to understand, if randomization can help achieve better bounds. 
For example, in Section \ref{approx_for_two_bidders}, we have proposed a truthful randomized mechanism for $2\times 2$ markets that improves on the approximation ratio of the deterministic golden ratio mechanism. 

Randomized approximation mechanisms, which have been designed for quasi-linear bidders do not necessarily carry over to markets with value bidders. Interestingly, \cite{Dobzinski12random} introduced a truthful and randomized approximation mechanism for quasi-linear bidders which obtains an optimal $O(\sqrt{m})$-approximation to the optimal revenue for arbitrary bidder valuations. This approximation ratio is tight, because it matches the best obtainable ratio for the algorithmic problem in polynomial time when truthfulness is not required. The approximation scheme is also truthful in a \textit{universal sense}. This form of incentive compatibility for randomized mechanisms is stronger than truthfulness \textit{in expectation}, which is used in other papers \citep{Lavi11}. Truthfulness in a universal sense requires that for any fixed outcome of the random choices made by the mechanism, players still maximize their utility by reporting their true valuations. In other words, a randomized mechanism is universally truthful if it is a probability distribution over truthful deterministic mechanisms.
It is therefore also incentive compatible for risk-averse bidders.

The general framework designed by \cite{Dobzinski12random} for mechanism design in quasi-linear settings, can easily be adapted for value bidders. Therefore, we can also achieve a randomized mechanism for combinatorial auctions with value bidders, which is truthful in a universal sense. 

The framework tries to distinguish two cases: either there is a dominant bidder such that allocating all items to him is a good approximation to the revenue, or there is no such bidder. In the first case all items will be assigned to a bidder. In the second case, a fixed-price auction is performed, which achieves a good approximation. In a first phase of the auction bidders are partitioned randomly in three sets. One of these sets is then used to gather statistics in phase II, which allow to set a reserve price in the second-price auction of phase III, which only allocates all items to one of the bidders. If the reserve price is not met in phase III, then the items are sold in the fixed-price auction in phase IV. 
 We will refer to this framework as $U$.

\begin{theorem} \label{thm-rand-gen-approx}
A randomized mechanism according to framework $U$ for value bidders is universally truthful and runs in polynomial time. It guarantees an approximation ratio of $O(\frac{\sqrt{m}}{\epsilon^3})$ with probability at least $1-\epsilon$.
\end{theorem}

\proof{} Except from the payment rule the framework $U$ is identical to the one described by \cite{Dobzinski12random}. Therefore, we can draw on the proof for their main theorem to show the approximation ratio of $O(\frac{\sqrt{m}}{\epsilon^3})$. Truthfulness in a universal sense is what remains to be shown. 
A randomized mechanism is truthful in the universal sense, if players still maximize their utility by reporting their true valuations for any fixed outcome of the random choices made by the mechanism. We show that the mechanism is truthful in each of the three partitions of bidders STAT, SECOND-PRICE, and FIXED over which the mechanism randomizes. 

Truthfulness in the SECOND-PRICE group is satisfied, because bidders can only win the entire bundle $J$ or nothing, which leads to an auction with single-parameter value bidders (see Theorem \ref{thm-single-parameter}). 
The fixed-price auction for the partition FIXED is truthful as well, since the bidder will try to win the most valuable bundle provided that his valuation is higher than the bundle price. Increasing the bid to a value higher than the reservation price in order to get a more valuable bundle, can result in a loss for the bidder because he has to pay more than his true valuation. Decreasing the bid just decreases the chance of winning, but would not increase the utility of a value bidder. 
Bidders in STAT never receive any items, and thus have no incentive to misreport their preferences. This holds for quasi-linear and for value bidders. 
\endproof

A randomized mechanism according to framework $U$ is also tight as it obtains an $O(\sqrt{m})$-approximation of the optimal revenue for general bidder valuations. This also shows, that there exists a gap between the power of randomized versus deterministic mechanisms. Whether such a gap exists for quasi-linear mechanism design is an open problem \citep{Lavi07}.

\section{Conclusions}
Quasilinear utility functions are widely used in auction theory, but real-world bidder preferences are often different. Only recently, deviations from pure payoff-maximization have been studied in the literature. While some of the theoretical work tries to relax the linearity assumption to allow for more general functions \citep{morimoto2015strategy, baisa2013auction}, we look at a specific utility function motivated by reports about industry practice in digital ad markets and other domains.  This function is characterized by value maximization subject to pre-defined budgets. It is interesting to understand if it is possible to find strategy-proof mechanisms for marketes with such value bidders. 

Interestingly, Pareto optimality is trivially satisfied, while strategy-proof mechanisms that maximize revenue are impossible except for markets with single-minded or single-valued value bidders. Approximation and randomization are often used as tools to achieve good revenue without giving up on truthfulness in quasi-linear mechanism design. We show that it is impossible to get an approximation ratio of more than $\frac{1}{n}$ for the general case with deterministic mechanisms. Only for restricted $2 \times 2$ multi-unit markets where the strongest bidder is commonly known, we propose the golden ratio mechanism with an approximation ratio of $\frac{\sqrt 5 - 1}{2}$. 

Randomization allows to close the gap between truthful approximation mechanisms for value bidders and approximation ratios for the purely algorithmic problem. Here, we can draw on prior work by \cite{Dobzinski12}, who considered auction design with quasi-linear utility functions. Interestingly, their framework can be applied for markets with value bidders as well. 

In this paper, we do not analyze such overall budget constraints across multiple items as in generalized assignment markets, where the bid language is restricted to price quotes for single objects.
Another avenue to explore is the restriction of the types of valuations of bidders, which might allow for higher approximation ratios than $\frac{1}{n}$ for the unrestricted case. While we have analyzed restricted cases with two units of a good and two bidders only, there might be other useful restrictions as well.

\bibliographystyle{model2-names}
\bibliography{literature}






\appendix
\section{Adaptation of the framework for randomized mechanisms by Dobzinksi et al.} \label{dobz-frame}
\DontPrintSemicolon
\begin{algorithm}[H]
\RestyleAlgo{plain}
\KwData{$n$ bidders, each with a general valuation $v_i$, a rational number $0 < \epsilon < 1$.}
\KwResult{An allocation of the items, which is an $O(\frac{\sqrt{m}}{\epsilon^3})$-approximation to the optimal revenue.}

\textbf{Phase I}: Partitioning the Bidders

\Indp 	Assign each bidder to exactly one of the following three sets: FIRST-PRICE with probability $1-\epsilon$, FIXED with probability $\frac{\epsilon}{2}$, and STAT with probability $\frac{\epsilon}{2}$.

\Indm	
\textbf{Phase II}: Gathering Statistics

\Indp 	Calculate the value of the optimal fractional solution in the combinatorial auction with all $J$ items, but only with bidders in STAT. Denote this value by $OPT^*_{STAT}$.

\Indm	
\textbf{Phase III}: A Second-Price Auction

\Indp 	Conduct a second-price auction with a reservation price of $r=\frac{OPT^*_{STAT}}{\sqrt{m}}$, in which the bundle $J$ of all items is sold to the bidders in SECOND-PRICE. If there is a ''winning bidder'', $i$, allocate all the items to that bidder, charge this bidder $\max(v_i(J), r)$, and output this allocation. Otherwise, proceed to the next step.

\Indm	


\textbf{Phase IV}: A Fixed-Price Auction

\Indp 	Let $R=J$. Let $p=\frac{\epsilon OPT^*_{STAT}}{8m}$.

For each bidder $i \in FIXED$, in some arbitrary order:
\begin{itemize}
	\item Let $S_i$ be the demand of bidder $i$ given the following prices: \\
	$p$ for each item in $R$, and $\infty$ for each item in $J - R$.
	\item Allocate $S_i$ to bidder $i$, and set his price to be $\max(v_i(S),  p \cdot |S_i|)$.
	\item Let $R = R \backslash S_i$.
\end{itemize}

\caption{Randomized approximation mechanism for general bidders}
\label{alg:dobzinski}
\end{algorithm}

\end{document}